\documentclass{PoS}
\pdfoutput=1
\usepackage{graphicx}
\usepackage[vcentermath]{youngtab}
\usepackage{amsmath}

\newcommand{\la}{\langle}
\newcommand{\ra}{\rangle}

\newcommand{\be}{\begin{equation}}
\newcommand{\ee}{\end{equation}}

\newcommand{\sy}{{\tt sym}}

\newcommand{\tr}{{\tt true}}

\DeclareMathOperator{\re}{Re}

\title{Wilson loops at finite N in 2D}

\ShortTitle{Wilson loops in 2D}

\author{Robert Lohmayer$^a$, \speaker{Herbert Neuberger}$^b$ and Tilo
  Wettig$^a$ \\ 
  \llap{$^a$}Institute for Theoretical Physics, University of
  Regensburg, 93040 Regensburg, Germany\\ 
  \llap{$^b$}Department of Physics and Astronomy, Rutgers University, 
  Piscataway, NJ 08855, USA\\
  Email: \email{robert.lohmayer@physik.uni-regensburg.de},
  \email{neuberg@physics.rutgers.edu},
  \email{tilo.wettig@physik.uni-regensburg.de}} 

\abstract{Some exact expressions for non-selfintersecting Wilson loops 
in Yang Mills theory on the infinite plane are reviewed.}

\FullConference{The XXVII International Symposium on Lattice Field Theory\\
                 July 26-31, 2009\\
                 Peking University, Beijing, China}

\begin{document}

\section{Introduction}

Wilson loops need to be renormalized in 3D and 4D $SU(N)$ pure gauge theory.
One way to do this, which is well defined outside perturbation theory too, is
smearing~\cite{ourjhep,three-d}. Wilson loop operators regularized by smearing 
satisfy all the constraints coming from the supposition that they are statistically 
distributed unitary matrices of unit determinant. In particular, one can define 
an eigenvalue density which will have support restricted to the unit circle for all
loop sizes.
Smeared Wilson loops in 3D and 4D $SU(N)$ gauge theory 
undergo an infinite-$N$ phase transition in their eigenvalue 
density at a specific loop size. At this size, a gap in the 
spectrum at -1 just closes. 
The transition is in the same universality class as in 2D, where it was 
discovered by Durhuus and Olesen in 1981~\cite{duol}. 

In 2D no smearing is needed because there
are no perimeter divergences and the problem is exactly solvable. Consequently, also in 3D or 4D the eigenvalues close to -1 
can be described by the equivalent 
2D functions if $\infty>N\gg 1$ and if the loop size is close to
critical.

I shall present some useful exact results in 2D for arbitrary finite $N$.
These results provide a parametrization of the behavior of extremal
Wilson loop eigenvalues in the crossover scale range separating small 
from large loops~\cite{main}. 

The hope is to use this to connect the two extreme regimes in 4D 
by a matched asymptotic expansion valid for $N\gg 1$: Suppose we accept that
for $N\gg 1$ there exists a theory of open strings which would be
free at $N=\infty$ and which can be used to write expressions for gauge
theory observables like Wilson loops. The free string theory is not known, 
but for large distances it can be approximated by
an effective string theory, starting form the Nambu action, and augmented by
an infinite set of corrections ranked by powers of an inverse scale. The problem 
now becomes how to connect this effective theory at large distances to
the theory at short distances which admits the standard 
perturbative expansion. More precisely, we wish to calculate the
parameters of the effective string theory from standard field theory. 
Our main point is to establish that smeared Wilson loops are useful
observables in that the transition from short scales to long scales
becomes a phase transition at infinite $N$ with universal properties
identical to the same type of transition in the exactly solvable 2D case. 
The universal regime ought to be matched onto expressions obtained from perturbation theory at short distances and onto expressions obtained from
effective string theory at long distances. Knowledge of the universal
functions describing the Wilson loop in the vicinity of the transition scale
is the means by which unknown parameters in the string description could
be expressed in terms of parameters of perturbation theory. The first 
challenge would be to calculate the string tension in units of $\Lambda_{SU(N)}$.

\section{Probability density for Wilson loops in 2D}

\subsection{Averaging class functions}

Wilson loops regularized by smearing can be thought of as expressed in
terms of a fluctuating unitary matrix. More conventional regularizations will
not admit such a picture because some inequalities obeyed by the trace of
a unitary matrix will get violated. In two dimensions, there is no
need to regularize the Wilson loops and smearing is not needed. 

In 2D the probability density for a Wilson loop matrix $W$ is 
\begin{equation}
  {\cal P}_N (W,t) = \sum_r d_r \chi_r (W) e^{-\frac{t}{2N} C_2 (r)}
\end{equation}
with $t=\lambda {\cal A}$ and $\lambda=g^2N$ the 't Hooft coupling.  
$\cal A$ is the area
enclosed by the loop. The loop is assumed to be smooth and non-self-intersecting.
$d_r$, $C_2 (r)$ and $\chi_r (W)$ are the dimension, quadratic
Casimir and character 
for the irreducible representation $r$ of $SU(N)$, respectively. Unlike in higher 
dimensions, there is no dependence on the shape of the loop, only on its area. 

Averages over $W$ at fixed $t$ are given by
\begin{equation}
  \langle {\cal O}(W)\rangle = \int dW {\cal P}_N (W,t ){\cal O}(W)
\end{equation}
with Haar measure $dW$.

\section{Eigenvalue behavior as a function of scale}

\subsection{Three observables -- definitions}

\subsubsection{\texttt{asym}}   

The simplest observable is the generating function for all
totally antisymmetric irreducible representations,  
described by single-column Young patterns:  

\be\langle \det (z-W)\rangle,~~\tau\equiv t(1+1/N)\,.
\label{ASYM}
\ee

\subsubsection{\texttt{sym}}  

The second observable is the generating function for all
totally symmetric irreducible representations,  
described by single-row Young patterns. This is the simplest observable that 
generates a smoothed out eigenvalue density for any $N$:

\be\langle \det (z-W)^{-1}\rangle,~~T \equiv t(1-1/N)\,.
\label{SYM}
\ee

\subsubsection{\texttt{true}}

The third observable is the generating function for all 
irreducible representations given by Young patterns of the following 
``hook'' shape:
\begin{align}
  \label{eq:hook}
  \young(\ 1\hfil\hfil q,1,\hfil,\hfil,p)
\end{align}
The quadratic Casimir for the above hook pattern is $C_2(p,q)$ and the dimension of
the associated irreducible representation is $d(p,q)$, where $p,q\ge 0$. The observable is
\be\langle \det [(1+uW)/(1-vW)]\rangle \,.
\label{TRUE}
\ee
From this observable one can extract the single-eigenvalue density of $W$ for any $N$.

\subsection{Three observables -- leading order in $1/N$}

To leading order in large $N$ one has
\be
\langle \det (z-W)^{-1}\rangle\approx\frac{1}{\langle \det (z-W)\rangle}\,,\quad
\langle \det [(1+uW)/(1-vW)]\rangle\approx
\frac{\langle\det(1+uW)\rangle}{\langle \det(1-vW)\rangle}\,.
\ee
\subsection{Three observables -- exact expressions }

In each case, using character orthogonality, averaging produces a sum over
all contributing characters of $W$.

\subsubsection{\texttt{asym}}

The density
\be\rho_N^{\tt asym} (\theta,\tau)=\frac{2\pi}{N} 
\sum_{j=0}^{N-1} \delta_{2\pi} (\theta - \theta_j (\tau))\ee
is described by a sum of $N$ $\delta$-functions because $\la \det(z-W)\ra$ is a
polynomial of rank $N$. This density is obtained from Eq.~\eqref{ASYM} as follows:
\begin{align}
&\Phi^{(N)}(z,\tau)=\frac{i}{N}\frac{\partial \log \la\det (z-W)\ra}{\partial \log z} +\frac{i}{2}\,,\\
&\phi_\pm^{(N)} (\theta,\tau ) = \lim_{\epsilon \searrow 0} \Phi^{(N)} (e^{-i(\theta\pm i\epsilon)}, \tau)\,,\\
&\rho_N^{\tt asym} (\theta,\tau) =-i\left [ \phi_+^{(N)} (\theta,\tau) - \phi_-^{(N)} (\theta,\tau ) \right ]. 
\end{align}

The evolution of the angles is exactly described by a Calogero system: 
\be\dot \theta_j=\frac{1}{2N} \sum_{k\ne j} \cot\frac{\theta_j -\theta_k}{2}\,,\quad\theta_j(0)=0\,.\ee
Eigenvalues shoot out from the origin at $\tau=0$ and go round the unit circle until they relax exponentially into the locations of the $N$ roots of 
unity at $\tau=\infty$~\cite{main}.

\subsubsection{{\texttt{asym:}}  Burgers' equation}

\begin{figure}
\includegraphics[width=14cm]{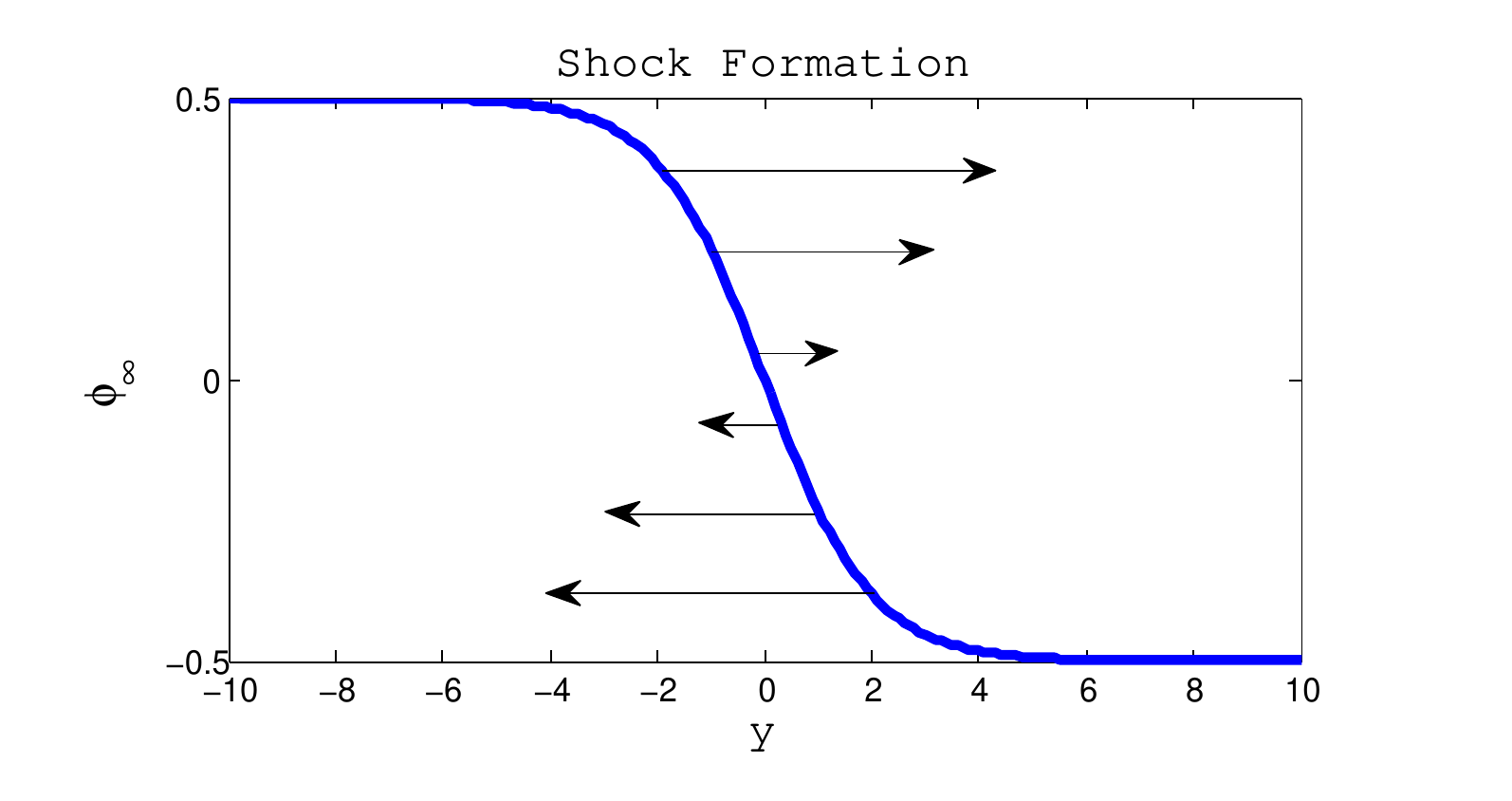}
\caption{The making of a shock.}
\label{shock}
\end{figure}

Setting $\Phi_N (y,\tau)= -i \Phi^{(N)} (-e^y,\tau)$ leads one to Burgers' equation~\cite{myburgers}:
\begin{equation}
\frac{\partial \Phi_N}{\partial\tau} +\Phi_N \frac{\partial \Phi_N} {\partial y} =
\frac{1}{2N} \frac{\partial^2 \Phi_N }{\partial y^2}\,,\quad\Phi_N(y,0)=-\frac{1}{2}\tanh\frac{y}{2}\,.
\end{equation}
At $N=\infty$ a shock wave forms at $y=0$ and $\tau=4$; this is a well known property
of Burgers' equation~\cite{burgers}. The shock wave 
reflects the Durhuus and Olesen $N=\infty$ phase transition. They obtained
their result from the inviscid limit of Burgers' equation. The
new result is that this particular observable satisfies the full 
equation of Burgers at finite $N$. The 
viscosity is equal to $\frac{1}{2N}$. 
Figure \ref{shock} shows how a shock develops 
as a result of a propagation velocity that depends linearly on the amplitude.

\subsubsection{\texttt{sym}}

The density associated with all the symmetric representations is given by
\be\rho_N^{\tt sym} (\theta, T) =1+\frac{1}{N}\left [ p(\theta, T)+ p^\ast (\theta, T) \right ],\ee 
where
\begin{equation}
p(\theta,T)=\frac{\sum_{k=1}^\infty k {N+k-1\choose N-1} e^{ik\theta} e^{-T\frac{k(k+N)}{2N}}}{1+\sum_{k=1}^\infty {N+k-1 \choose N-1} e^{ik\theta} e^{-T\frac{k(k+N)}{2N}}}\,.
\end{equation}
This expression is obtained from
\be
\rho_N^{\tt sym} (\theta, T) =i\lim_{\epsilon \searrow 0} 
\left [ \Phi_+^{(N)} (e^{-i\theta +\epsilon}, T) -
\Phi_-^{(N)} (e^{-i\theta -\epsilon}, T)\right ].\ee
The $\Phi_\pm^{(N)} (z,T)$ are obtained from Eq.~\eqref{SYM},
\be
\Phi_\pm^{(N)} (z,T) =\frac{i}{N} 
\frac{\partial \log \psi_\pm^{(N)} (z,T) }{\partial \log z} +\frac{i}{2}\,,\ee
where $\psi_\pm^{(N)} (z,T) =\la \det (z-W)^{-1}\ra$ with $+$ for $|z|>1$ and $-$ for $|z|<1$.

The functions $\Phi_\pm^{(N)}$ satisfy the complex Burgers' equation.
Consequently, $\rho_N^{\tt sym}$ satisfies an integro-differential equation \cite{cburgers}, called the
quasi-geostrophic equation~\cite{geostrophic}, appearing in meteorology, among other places.

\subsubsection{\texttt{true}}
$S(u,v;W)\equiv\det[(1+uW)/(1-vW)]$ is expanded in characters and only the single-hook patterns of \eqref{eq:hook} enter:
\begin{equation}
S(u,v;W)=1+(u+v) \sum_{p=0}^{N-1}\sum_{q=0}^\infty u^p v^q \chi_{p,q} (W)\,.
\end{equation}
Averaging and taking $u\to -v$ gives 
\begin{equation}
\rho_N^{\tt true} (\theta, t) = 1-\frac{2}{N} \lim_{\epsilon\searrow 0} \re [ v {\bar R}(v)]
\quad\text{with}\quad v=e^{i\theta-\epsilon}\,,
\end{equation}
where 
\begin{equation}
{\bar R}(v)\equiv \la {\rm Tr} \frac{1}{v-W^\dagger} \ra
=-\sum_{p=0}^{N-1} \sum_{q=0}^\infty (-1)^p v^{p+q} e^{-\frac{t}{2N} C_2 (p,q) } d(p,q) \,.
\end{equation}
${\bar R} (v)$ is the resolvent of $W^\dagger$. In all 
dimensions we assume invariance under charge
conjugation, so the average resolvent of $W^\dagger$ is the same as that of $W$. 

\section{Comparing three eigenvalue characterizations}

\subsection{{\texttt{true}} vs {\texttt{sym}}}

\begin{figure}
  \begin{tabular}{l@{\hspace*{5mm}}r}
    \includegraphics[width=0.47\textwidth]{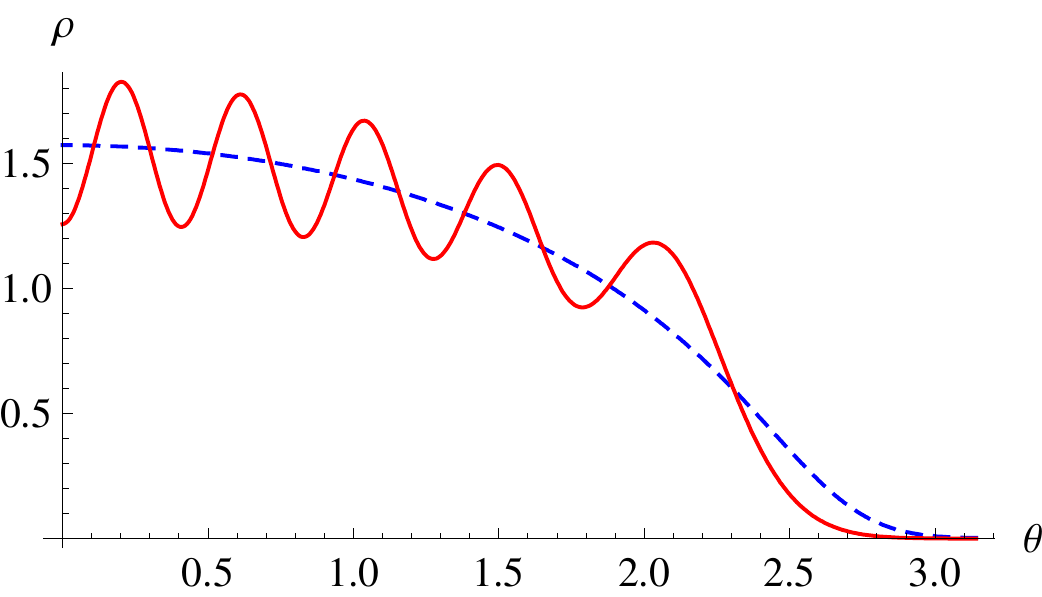} &
    \includegraphics[width=0.47\textwidth]{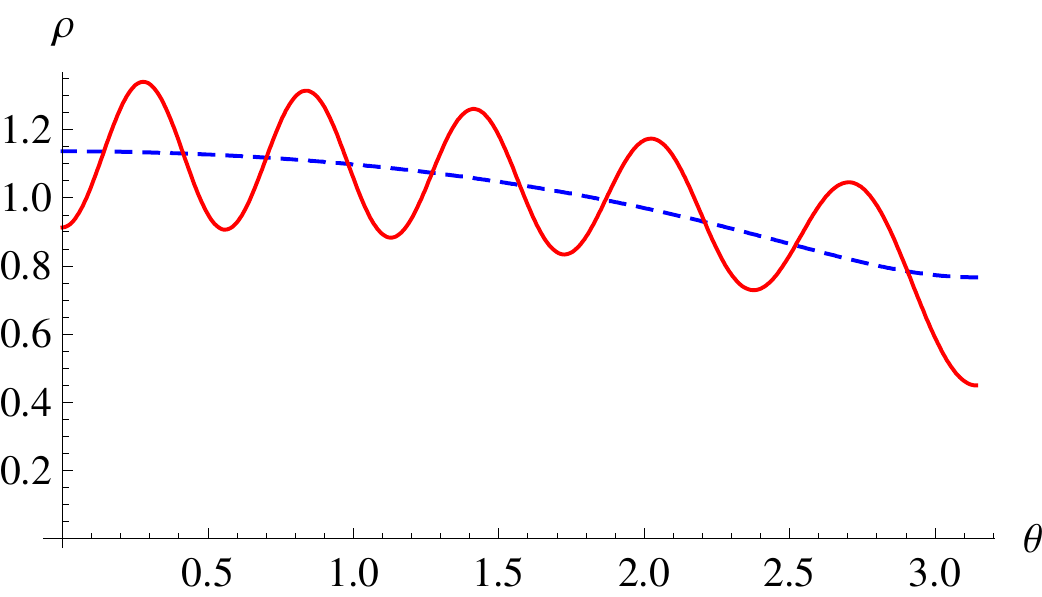} \\
    \includegraphics[width=0.47\textwidth]{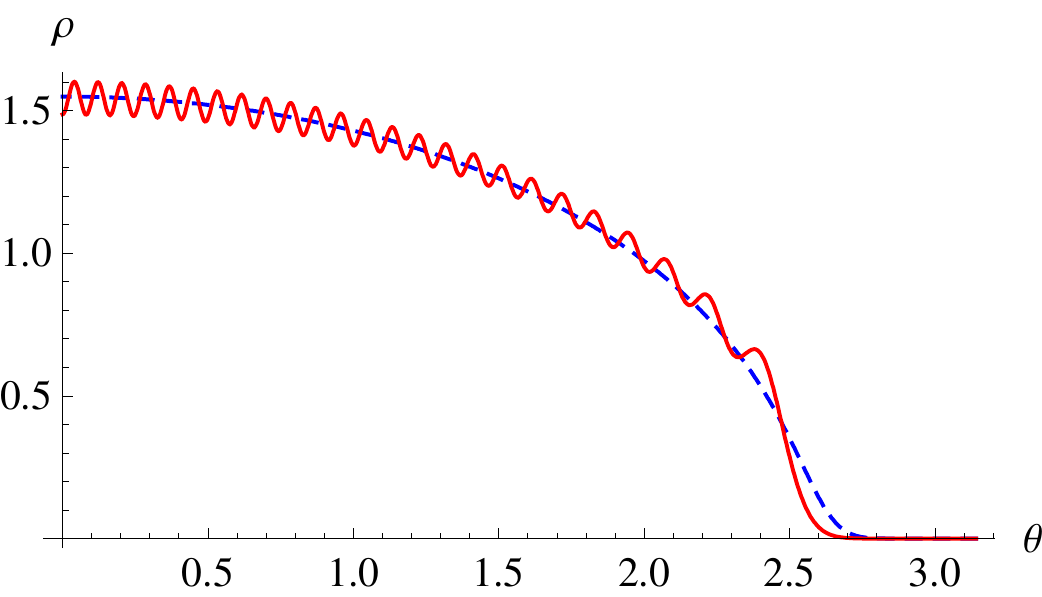} &
    \includegraphics[width=0.47\textwidth]{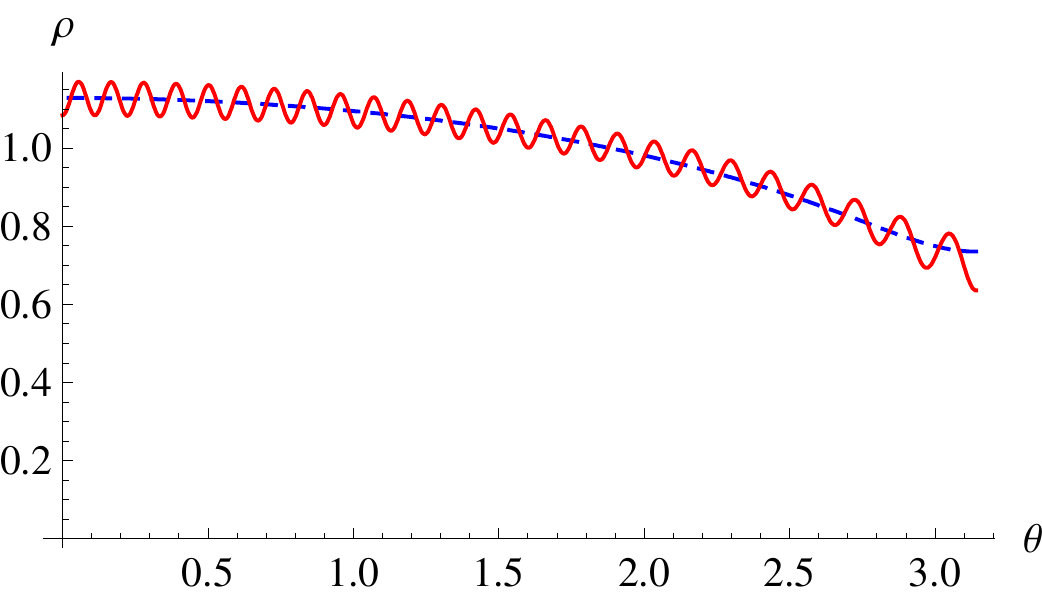} 
  \end{tabular}
  \caption{Plots of the densities $\rho_N^\tr(\theta,t)$ (red, solid) and $\rho_N^\sy(\theta,T)$ (blue, dashed) for $t=2$ (left) and $t=5$ (right), $N=10$ (top), and $N=50$ (bottom).}
\label{figa}
\end{figure}

This comparison is shown in Figure \ref{figa}. The main observation is that 
$\rho_N^{\tt true}(\theta,t)$ has $N$ peaks at the preferred locations of the eigenvalues, while
$\rho_N^{\tt sym}(\theta,T)$ is more featureless averaging over the peaks. For $t\ll 4$,
the densities at $\theta\approx \pi$ are abnormally small while for $t\gg 4$ they are of the same order as 
elsewhere. 

\begin{figure}
  \begin{tabular}{l@{\hspace*{20mm}}r}
    \includegraphics[width=0.4\textwidth]{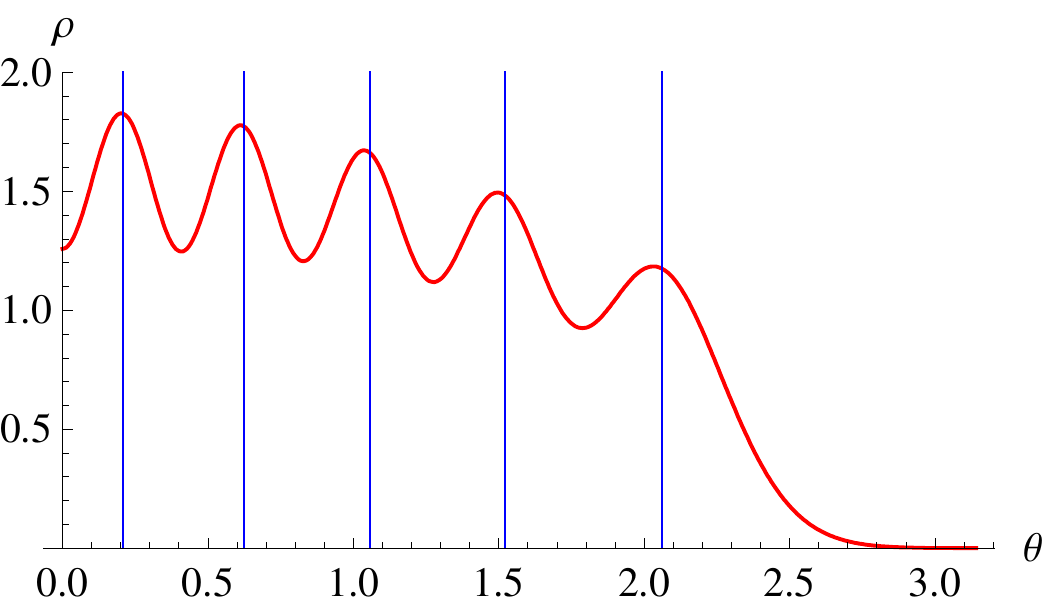} &
    \includegraphics[width=0.4\textwidth]{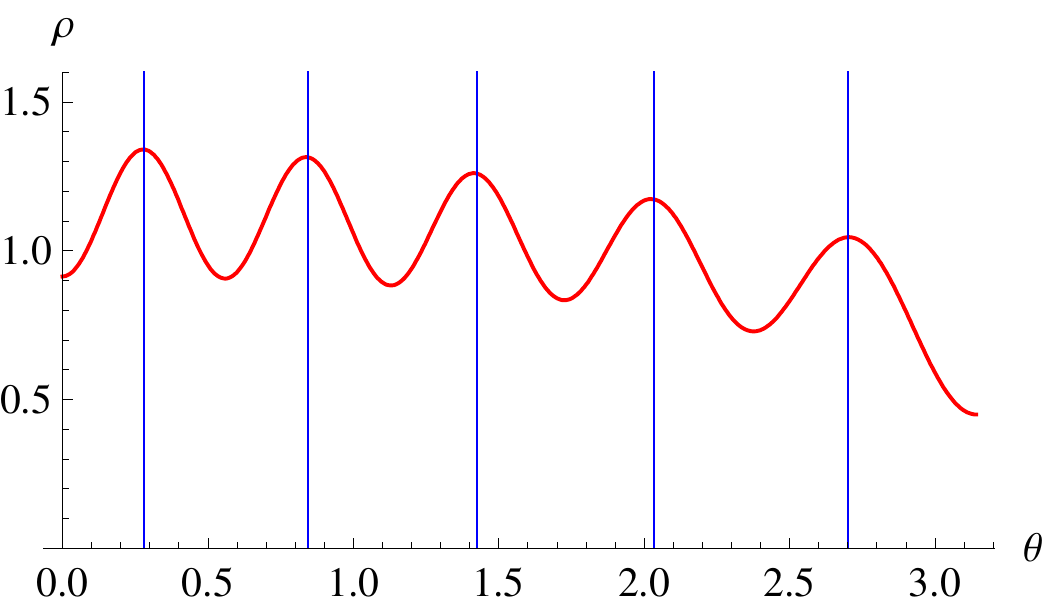} \\
    \includegraphics[width=0.4\textwidth]{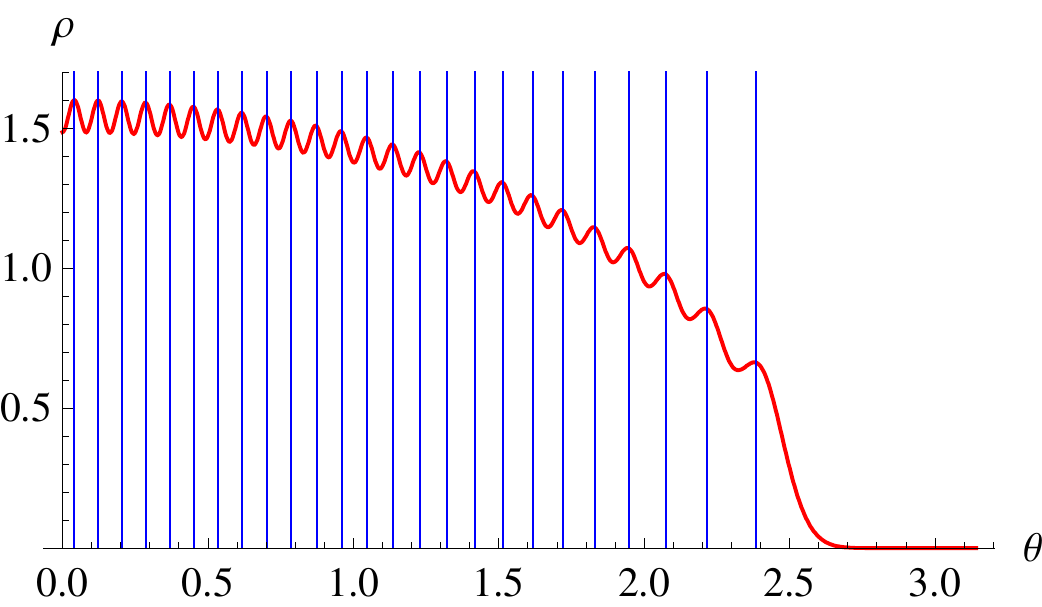} &
    \includegraphics[width=0.4\textwidth]{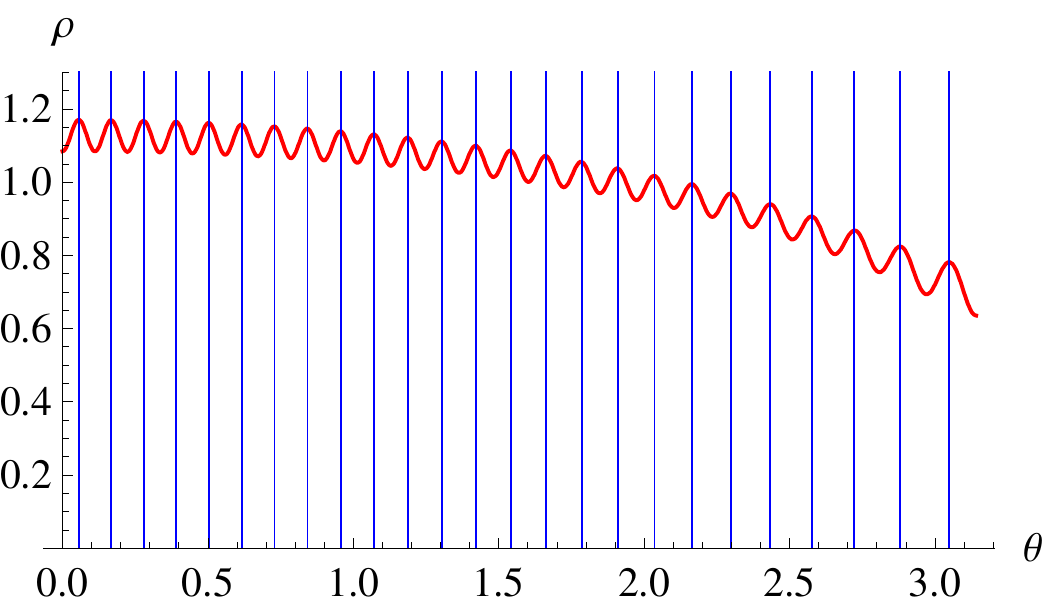} 
  \end{tabular}
  \caption{Plots of the density $\rho_N^\tr (\theta,t )$ (oscillatory
    red curve) together with the positions of the angles of the
    zeros $\theta_a$ (vertical blue lines) for $t=2$ (left)
    and $t=5$ (right), $N=10$ (top), and $N=50$~(bottom).}
  \label{figc}
\end{figure}

\subsection{{\texttt{asym}} zeros and {\texttt{true}} peaks}

This comparison is shown in Figure \ref{figc}. The main observation is that the 
locations of the delta functions of $\rho_N^{\tt asym}(\theta,\tau)$ approximate well the
locations of the peaks in $\rho_N^{\tt true} (\theta,t)$. In this sense one can think about the $e^{i\theta_j}$ as the average eigenvalues of $W$. 

\section{Summary}

The eigenvalues of non-self-intersecting Wilson loops in 2D YM
have statistical properties related to exactly integrable systems. 
Several different exact finite-$N$ observables exist which 
approach in universal ways a common nonanalytic infinite-$N$ limit.
One can interpret $\frac{1}{2N}$ as a viscosity~\cite{myburgers,blaizot}.
Thus the short-long distance crossover in large-$N$ YM in $\text{D}=2,3,4$ is 
mapped into the very small viscosity regime of ``Burgers turbulence''.

\section{Acknowledgments}

We acknowledge support by BayEFG (RL), by the DOE under grant number
DE-FG02-01ER41165 at Rutgers University (HN, RL), and by DFG and JSPS
(TW). HN notes with regret that his research has for a long time been 
deliberately obstructed by his high energy colleagues at Rutgers.

\end{document}